\documentclass[preprint,nohyper,notoc]{JHEP3}
\usepackage{graphicx}
\usepackage{amsmath}
\usepackage{epsf}
\usepackage{psfrag}
\usepackage[sort&compress,numbers]{natbib}

\def\be{\begin{equation}}
\def\ee{\end{equation}}
\def\bea{\begin{eqnarray}}
\def\eea{\end{eqnarray}}

\def\e{\epsilon}

\def\la{\langle}
\def\ra{\rangle}

\def\Psl{\not{\hbox{\kern-2.3pt $P$}}}
\def\psl{\not{\hbox{\kern-2.3pt $p$}}}
\def\qsl{\not{\hbox{\kern-2.3pt $q$}}}
\def\Ksl{\not{\hbox{\kern-2.3pt $K$}}}
\def\ksl{\not{\hbox{\kern-2.3pt $k$}}}
\def\esl{\not{\hbox{\kern-2.3pt $\pol$}}}

\def\pol{\varepsilon}

\def\spa#1.#2{\left\langle#1\,#2\right\rangle}
\def\spb#1.#2{\left[#1\,#2\right]}
\def\lor#1.#2{\left(#1\,#2\right)}
\def\sand#1.#2.#3{%
\left\langle\smash{#1}{\vphantom1}^{-}\right|{#2}%
\left|\smash{#3}{\vphantom1}^{-}\right\rangle}
\def\sandp#1.#2.#3{%
\left\langle\smash{#1}{\vphantom1}^{-}\right|{#2}%
\left|\smash{#3}{\vphantom1}^{+}\right\rangle}
\def\sandpp#1.#2.#3{%
\left\langle\smash{#1}{\vphantom1}^{+}\right|{#2}%
\left|\smash{#3}{\vphantom1}^{+}\right\rangle}
\def\sandpm#1.#2.#3{%
\left\langle\smash{#1}{\vphantom1}^{+}\right|{#2}%
\left|\smash{#3}{\vphantom1}^{-}\right\rangle}
\def\sandmp#1.#2.#3{%
\left\langle\smash{#1}{\vphantom1}^{-}\right|{#2}%
\left|\smash{#3}{\vphantom1}^{+}\right\rangle}
\def\sandmm#1.#2.#3{%
\left\langle\smash{#1}{\vphantom1}^{-}\right|{\slash\!\!\! #2}%
\left|\smash{#3}{\vphantom1}^{-}\right\rangle}
\def\spab#1.#2.#3{\sandmm#1.#2.#3}
\def\spbb#1.#2.#3.#4{\sandpm#1.{\slash\!\!\! #2\slash\!\!\! #3}.#4}
\newbox\charbox
\newbox\slabox
\def\s#1{{      % Feynman slash
        \setbox\charbox=\hbox{$#1$}
        \setbox\slabox=\hbox{$/$}
        \dimen\charbox=\ht\slabox
        \advance\dimen\charbox by -\dp\slabox
        \advance\dimen\charbox by -\ht\charbox
        \advance\dimen\charbox by \dp\charbox
        \divide\dimen\charbox by 2
        \raise-\dimen\charbox\hbox to \wd\charbox{\hss/\hss}
        \llap{$#1$}
}}
\def\ksl{\s{k}}

\def\beqa{\begin{eqnarray}}
\def\eeqa{\end{eqnarray}}
\def\beq{\begin{equation}}
\def\eeq{\end{equation}}

\newcommand{\wh}[1]{\widehat{#1}}

\def\ket#1{|#1\rangle}
\def\BB#1#2#3{[#1|#2|#3\rangle}
\def\AA#1#2#3{\langle#1|#2|#3]}
\def\Aa#1#2#3{\langle#1|#2|#3\rangle}
\def\Bb#1#2#3{[#1|#2|#3]}
\def\A#1#2{\langle#1#2\rangle}
\def\B#1#2{[#1#2]}

\preprint{
  hep-th/0507161\\
  IPPP/05/39\\
  DCPT/05/78\\
  July, 2005}

\title{Recursion Relations for Gauge Theory Amplitudes with Massive Vector Bosons and Fermions}

\author{S. D. Badger,
    \ E. W. N. Glover, Valentin V. Khoze \
    \\
        Department of Physics,
        University of Durham, Durham, DH1 3LE, UK\\
        E-mails: \email{s.d.badger@durham.ac.uk,
        e.w.n.glover@durham.ac.uk,
        valya.khoze@durham.ac.uk}.
    }

\abstract{
We apply the on-shell
tree-level recursion relations of Britto, Cachazo, Feng and Witten
to a variety of processes
involving internal and external massive particles with spin.
We show how to construct multi-vector boson currents where one or more 
off-shell vector bosons couples to a quark pair and number of gluons.
We give compact results for single vector boson currents with up to six partons 
and double vector boson currents with up to four partons for all
helicity combinations.   We also provide expressions for single vector boson currents
with a quark pair and an arbitrary number of gluons for some specific helicity configurations.
Finally, we show how to generalise the recursion relations to handle 
massive particles with spin on internal lines using $gg \to t\bar t$ as an example.}

\begin{document}
\bibliographystyle{JHEP-2}
\section{Introduction}

Powerful new formalisms have been recently developed for the calculation of on-shell
quantities such as scattering amplitudes in gauge theories.   Most striking are the
MHV rules of Ref.~\cite{CSW:MHVtree} and the Britto-Cachazo-Feng-Witten (BCFW)
recursion relations of Refs.~\cite{BCF:rec,BCFW:proof}. Applications of these new
formalisms together with the classic unitarity based approach of
Refs.~\cite{BDDK:unitarity1,BDDK:unitarity2} have led to a dramatic progress in
calculations of amplitudes.

The `MHV rules', originally proposed in \cite{CSW:MHVtree}, led to simple and compact
calculations of tree level amplitudes
\cite{Zhu:googly1,GK:MHVtrees,Kosower:NMHV,Wu:MHV1,Wu:MHV2,GGK:nonMHV,Khoze:review}.
The method has also been generalised to include external Higgs bosons and massive
vector bosons \cite{Dixon:MHVhiggs,Badger:MHVhiggs2,Bern:EWcurrents} as well for
investigating collinear factorisation \cite{Birthwright:MHVcoll,Birthwright:MHVcoll2}.

The on-shell BCFW tree-level recursion relations were first formulated  for massless
particles~\cite{Roiban:2004ix,BCF:rec,BCFW:proof,Luo:recfermions,Luo:cmpct6pt,Britto:2005dg}. 
The proof of the new recursion relations \cite{BCFW:proof} relies on simple analytic
properties of the amplitudes and have recently been generalised in three ways. First,
the massless on-shell recursion relations in gauge theory  were extended to include
gravity~\cite{Bedford:recgenrel,Cachazo:recgenrel}. Second,  a new version of
recursion relations was developed in  Refs.~\cite{BDK:1lonshell,BDK:1lrecfin} to
calculate all finite one-loop amplitudes in non-supersymmetric QCD. 
Ref.~\cite{Bern:bootstrap} further generalised this approach to compute the
cut-nonconstructible parts of divergent one-loop gluonic amplitudes. In a third
development,  Ref.~\cite{mrscalar} generalised the BCFW recursion relations to include
massive scalar particles at tree level. The goal of this paper is to further extend
the recursion relations to include massive particles with spin such as vector bosons
or massive quarks in a natural way. We will show that the recursion relations lead to
new compact formulae for scattering amplitudes that are somewhat simpler than the
previous  calculations of Berends, Giele and
Kuijf~\cite{Berends:recurs1,Berends:vector}.

Much progress has also been made at one loop level. The MHV rules have been
successfully applied to compute supersymmetric amplitudes
\cite{CSW:1l,BST:1lMHV-N4,CSW:holomorphicanomoly,Bena:loops,Quigley:1lMHV,BBST:1lMHV-N1,BBST:1lMHV-nonsusy}
as well as new calculations using improved unitarity methods
\cite{Cachazo:HA+uni,BCF:1lHA,Bern:NMHV1l7pt,Bidder:1lHA-N1,BCF:gu,BDK:allNMHV1l,
Bidder:1lN<4,Britto:sQCD,Brandhuber:ddimgu, Buchbinder:2luni,Bern:coeffsrec}. The new
formalism has been largely inspired by Witten's idea of a duality between
supersymmetric Yang-Mills and a topological string theory on a twistor
space~\cite{Witten:twstr}. For reviews of recent developments see
\cite{Cachazo:twstrreview,Nair:twistornotes}.

Our paper is organised as follows. In section~{\bf \ref{sec:rev}} we review the
recursive formulation of gauge theory amplitudes. Section~{\bf \ref{sec:model}}
defines our model and reviews colour ordering of the amplitudes.  We provide new
compact results for amplitudes with up to six partons and a single vector boson in
section~{\bf \ref{sec:single}} and with up to four partons and two vector bosons in
section~{\bf \ref{sec:double}}. New compact expressions for $n$-point NMHV currents with
a single vector boson are also given in section~{\bf \ref{sec:single}}. Finally we
demonstrate how to use recursion relations including propagating massive particles
with spin in section~{\bf \ref{sec:spin}}.  Our findings are briefly summarized in
section~{\bf \ref{sec:conc}}.   An appendix stating our spinor conventions is also
included.
\newpage

\section{On-Shell Recursion Relations \label{sec:rev}}

Here we give a brief summary of the on-shell recursion relations \cite{BCF:rec} for
tree-level amplitudes. These recursion relations follow from general quantum field
theory arguments \cite{BCFW:proof} and we will present them in the form
\cite{mrscalar} which naturally incorporates massless and massive particles. We
consider the colour-ordered partial amplitudes ${\cal A}={\cal A} (p_1,\dots, p_n),$
in which the coloured particles come in a definite cyclic order $1,2,\dots, n$. These
amplitudes are obtained by stripping away the colour factors from the full amplitude,
hence, they depend on the kinematic variables, momenta and helicities, $p_k$ and $h_k$
only.

The formalism relies on choosing two particles in the amplitudes to be shifted by a
complex vector to be specified below.
We label these marked particles $i$ and $j$.
The on-shell recursion relation for a tree-level amplitude takes the form:
\begin{align}
\mathcal{A}_n (p_1,\ldots,p_n) =\,
    \sum_{{\rm partitions}}\sum_{h}
        & \mathcal{A}_L(p_r,\ldots,\wh{p}_i,\ldots,p_s,-\wh{P}^h) \,
    \frac{1}{P^2-m_P^2} \nonumber \\
    &\times \mathcal{A}_R(\wh{P}^{-h},p_{s+1},\ldots,\wh{p}_j,\ldots,p_{r-1}) \ ,
    \label{eq:rrel}
\end{align}
where $P=p_r+\ldots+p_i+\ldots+p_s$ and the
``hatted'' quantities are the shifted on-shell momenta.
The summation in \eqref{eq:rrel} is over all partitions of $n$ external particles
between the smaller amplitudes on the left, ${\cal A}_L$,
and on the right, ${\cal A}_R$, such that $p_i$ is on the
left, and $p_j$ is on the right, and also over the helicities,
$h,$ of the intermediate state.

In this paper we will always choose both marked particles $i$ and $j$
to be massless. We shift two massless momenta\footnote{Helicity spinors 
$\ket{{i}}$,  $|{i}]$ are defined in the Appendix
along with our conventions for forming the inner product.}
${p}_i=\ket{{i}}|{i}]$ and ${p}_j=\ket{{j}}|{j}]$ of the marked particles by
 $\eta=\ket{j}|i]$, such that the shifted momenta are
\bea \label{eq:225}
 \widehat{p}_i &=&  p_i + z |j\rangle|i]\ , \qquad \widehat{p}_i^2 = 0 = p_i^2\ , \\
\label{eq:226}
\widehat{p}_j &=&  p_j - z |j\rangle|i] \ ,
 \qquad \widehat{p}_j^2 = 0 = p_j^2 \ , \\
\label{eq:227}
\widehat{P} &=& P + z |j\rangle|i] \ , \qquad \widehat{P}^2 = m_P^2
 \ .
\eea
Here $m_P$ is the mass of the particle on the internal line.
Equation \eqref{eq:227} determines the variable $z$ as
\be
 \label{eq:228}
 z =\,  -\, {{P}^2-m_p^2\over 2 {P} \cdot \eta} =\,
-\frac{P^2-m_p^2}{ \AA{j}{P}{i} } \ .
\ee
For the particles $i,j$ momentum shifts above are equivalent
to shifting the spinors as follows:
\begin{eqnarray}
&&|\widehat{i}] = |i]  \ , \qquad \, \, |\widehat{i}\rangle = |i\rangle
+ z |j\rangle \ ,
    \label{eq:shiftsi} \\
    &&|\widehat{j}\rangle = |j\rangle \ , \qquad
    |\widehat{j}] = |j]- z |i] \ .
    \label{eq:shiftsj}
\end{eqnarray}

A direct proof of \eqref{eq:rrel} was given in reference \cite{BCFW:proof}
based on the analyticity structure of the meromorphic tree-level amplitude
$\mathcal{A}_n(z)$ in the complex $z$-plane. For equation
\eqref{eq:rrel} to hold it is essential that $\mathcal{A}_n(z)$ should have no
poles at infinity. In general it is known that whether or not these ``boundary''
contributions are present (i.e. whether or not \eqref{eq:rrel} is valid)
can depend on the choice of the marked particles $i$ and $j$
\cite{BCFW:proof,Luo:cmpct6pt,mrscalar}. For the classes of models considered in
this paper we will be always marking only massless particles, and in such a way
that the helicities of the
marked particles can take the values,
\be \label{hconds}
 (h_i,h_j)\,
=\,(+,-)\ ,\ (+,+)\ , \ (-,-)
\ee
but not $(h_i,h_j)=(-,+).$ In addition to this, if the two marked particles are
a quark and an anti-quark of the same flavour, they should not be adjacent. 
Finally, for adjacent quarks and gluons of equal helicity cases we make the choices
$(j_q^+,i_g^+)$ or $(i_q^-,j_g^-)$.

%%%%%%%%%%%%%%%%%%%%%%%%%%%%%%%%%%%%%%%%%%%%%%%%%%%%%%%%%%%%%%%%%

\section{Massive Vector Bosons \label{sec:model}}

The principal goal of this paper is to apply on-shell recursion relations to tree-level
amplitudes involving massive or off-shell particles with spin coupled to massless particles.
To begin with we will consider massive vector bosons coupled to massless
gauge fields and fermions.
Essentially we consider a generic theory with a non-Abelian gauge group being a product
$G_1 \times G_2$, where $G_1$ is unbroken, and $G_2$ is broken by the Higgs mechanism.
Gauge fields of the $G_1$ group are massless and the gauge fields of broken group $G_2$ are
massive or off-shell vector bosons $V.$
Two gauge groups are coupled to each other via fermions which are charged under both groups.
We will use the `colour decomposition' representation for scattering amplitudes with respect to
both groups, and hence the colour-stripped amplitudes will be purely kinematic quantities
which will not depend on the choice of $G_1$ and $G_2$ nor on the representations for the matter
fermions.

This set-up is rather general, and in particular it incorporates the elements of the Standard Model.
In this case $G_1$ is $SU(3)$ and the corresponding gauge fields are gluons $g$;
the gauge fields of the (partially) broken group, $G_2=SU(2)\times U(1),$ are
massive or off-shell vector bosons
$W^{\pm},$ $Z^0$ and $\gamma^*$. The fermions can be taken to be (anti)-quarks, $\bar{q}$, $q,$
transforming in the (anti)-fundamental representations of both groups.
Even in the general case, we will continue denoting
massless gauge fields as gluons, and fermions as quarks. 
Massive vector bosons will be denoted as $V$'s.

The quantities we want to consider are the $G_1$- and $G_2$-colour-stripped
purely kinematic tree-level amplitudes
\begin{equation}
    S_{\mu_1 \ldots \mu_m}(1_q,2,3,\ldots,n-1,n_{\bar{q}}).
    \label{eq:Sdefdcs}
\end{equation}
These are the $(m+n)$-point amplitudes with $m$ massive vector bosons
$V_{\mu_1},\ldots V_{\mu_m}$ coupled to $n$ massless partons. More specifically,
we consider the case of a single quark-antiquark pair\footnote{One cannot have less
than one $q\bar{q}$-pair in amplitudes coupling $V$'s to $g$'s at tree level.
Amplitudes with more than one $q\bar{q}$-pair will not be considered in this paper,
but they can be calculated in a similar way.}
denoted in
\eqref{eq:Sdefdcs}
as $1_q, n_{\bar{q}}$
and $n-2$ gluons labelled $2,3,\ldots,n-2.$

The group-theoretical dependence in the amplitudes can be easily restored in the usual way.
For the case of fundamental fermions the amplitude \eqref{eq:Sdefdcs} is multiplied by
$(T^{a_2}\ldots T^{a_{n-1}})_{i_1 i_n}$ and by
$(T^{b_1} \ldots T^{b_m})_{k_q k_{\bar{q}}}$. Then it is summed over all permutations of
$a_2, \ldots, a_{n-1}$ and over all permutations of $b_1, \ldots, b_m.$ Here
$T^a$ and $T^b$ are the generators of the $G_1$ and the $G_2$ groups respectively.

The physical states corresponding to all massless particles in amplitudes
\eqref{eq:Sdefdcs} will always be represented in the helicity basis, e.g.
$1_q^-,2^+,3^-\ldots,(n-1)^+,n_{\bar{q}^-}.$
At the same time, for the massive (off-shell) vectors $V_{\mu_1},\ldots V_{\mu_m}$ we
will always choose to not multiply them by external wave functions,
and instead of helicities or polarisations they will be characterised
by their Lorentz indices ${\mu_1 ,\ldots, \mu_m}.$
Thus, the amplitudes $S_{\mu_1 \ldots \mu_m}$ are the multi-vector boson currents.

Working with multi-currents \eqref{eq:Sdefdcs} will first of all facilitate our calculation:
single vector currents will be used in calculations of double currents and so on,
as will be seen in section {\bf \ref{sec:double}}.
Furthermore, multi-currents can be easily used to calculate full physical amplitudes
which include the
decay of the massive (off-shell) vector bosons into light stable states.
This is achieved by contracting
each Lorentz index $\mu$ in \eqref{eq:Sdefdcs} with the current $L^\mu$ describing
the relevant decay mode of each vector boson $V_\mu$.

In the Standard Model, for example, one can consider decays of unstable vector bosons
into a fermion-antifermion (lepton or quark) pair, so that
for virtual photon $\gamma^*$ or for
$V=W^\pm,Z$ boson decay we have:
\begin{align}
    L^\mu_{\gamma^*} &= -e^2 Q_f Q_q\frac{\bar{u}(p_f)\gamma_\mu u(p_{\bar{f}})}
    {P_{\gamma^*}^2} \nonumber\\
    L^\mu_{V} &= -e^2 \frac{v_{V;H}^{f\bar{f}} v_{V;H}^{q\bar{q}} \bar{u}(p_f)\gamma_\mu
    u(p_{\bar{f}})}{(P_{V}^2-M_{V}^2+i\Gamma_{V}M_{V})}.
    \label{eq:fermioncurrent}
\end{align}
Here the couplings $v_{V;H}$ for $V$ either $W$ or $Z$ bosons with either
left (L) or right (R) handed  polarisations are given by
\def\stw{\sin{\theta_w}}
\def\ctw{\cos{\theta_w}}
\begin{align}
    \begin{array}{ccc}
        v_{Z;R}^{f\bar{f}} = -Q_f\frac{\stw}{\ctw} \ ,&\quad
        v_{Z;L}^{f\bar{f}} = -Q_f\frac{I^3_{w,f}-\sin^2\theta_w Q_f}{\stw\ctw} \ , &\quad
        v_{W;L}^{\nu_i\bar{l}_j} = \frac{1}{\sqrt{2}\stw} \delta_{ij} \ , \\\\
        v_{W;L}^{l_i\bar{\nu}_j} = \frac{1}{\sqrt{2}\stw} \delta_{ij}\ , &
        v_{W;L}^{u_i\bar{d}_j} = \frac{1}{\sqrt{2}\stw} U_{ji}^\dagger \ , &
        v_{W;L}^{d_i\bar{u}_j} = \frac{1}{\sqrt{2}\stw} U_{ji} \ .
    \end{array}
    \label{eq:ewcouplings}
\end{align}
and all others zero. $U_{ij}$ is the CKM mixing matrix and the rest of notation is standard.

%%%%%%%%%%%%%%%%%%%%%%%%%%%%%%

\section{Single Vector Boson Currents \label{sec:single}}

The single vector boson currents were previously calculated by Berends, Giele and Kuijf
\cite{Berends:vector} using the recursive technique based on
iterations of classical equations of motion~\cite{Berends:recurs1}.
 More recently these single vector currents
were also discussed and derived in \cite{Bern:EWcurrents} using a combination of
Berends-Giele recursion relations and the MHV rules of \cite{CSW:MHVtree}.
Here we will employ the BCFW on-shell
recursion relations of section {\bf \ref{sec:rev}} to derive slightly more
compact expressions as well as  new results for
$n$-parton single currents for some specific helicity arrangements of partons.
As mentioned earlier, single off-shell currents $S_\mu$ are not only interesting
on their own right,
more significantly, they play an important part in a recursive construction
of currents with
two and more vector bosons as will be explained in the next section.
We note that a similar observation
has also been made in \cite{Bern:EWcurrents}.

We now proceed to construct the single vector boson currents,
\be
S_\mu(1_q^\lambda,2^{h_2},\ldots,(n-1)^{h_{n-1}},n_{\bar{q}}^{-\lambda}).
\ee
The on-shell recursion relations construct amplitudes from
on-shell amplitudes with fewer particles.
Assuming that one can always avoid contributions at $z \to \infty$ and using the
recursion relation $n-3$ times gives a representation of the
$n$-point amplitude entirely in terms of the 3-point vertices.
Hence, $3$-point vertices are the building blocks of all larger amplitudes and hence
the on-shell recursion relations reduce the task of computing general
amplitudes to the computation of all 3-point vertices.
This is indeed the case  in theories with massless vectors coupled to fermions
and also to massless and massive scalars\footnote{This is however
not always the case in theories involving
scalar self-interactions. The 4-point vertex
corresponding to a $\phi^4$ interaction clearly cannot be reduced to 3-point vertices.
Recursion relation cannot be applied to this vertex
since it is a constant. Hence the corresponding amplitude $A_4(z)$ 
does not depend on $z$, and this
necessarily leads to a non-vanishing contribution at $z \to \infty$.
We thank George Georgiou for pointing this out to us.}
\cite{BCF:rec,BCFW:proof,mrscalar}.

The three-gluon and the quark-gluon-antiquark
primitive amplitudes are given by the standard MHV and
$\overline{\rm MHV}$ expressions:
\bea
\mathcal{A}_3(1^-,2^{-},3^+) &= \,
\frac{{\A{1}{2}}^3}{\A{2}{3}\A{3}{1}}\ , \qquad
\mathcal{A}_3(1^+,2^{+},3^-) &= \,
-\frac{{\B{1}{2}}^3}{\B{2}{3}\B{3}{1}}\ , \\
\mathcal{A}_3(1_q^-,2^{-},3_{\bar{q}}^+) &= \,
\frac{{\A{1}{2}}^2}{\A{1}{3}}\ , \qquad
\mathcal{A}_3(1_q^-,2^{+},3_{\bar{q}}^+) &= \,
-\frac{{\B{2}{3}}^2}{\B{1}{3}}\ , \\
\mathcal{A}_3(1_q^+,2^{-},3_{\bar{q}}^-) &= \,
\frac{{\A{2}{3}}^2}{\A{1}{3}}\ , \qquad
\mathcal{A}_3(1_q^+,2^{+},3_{\bar{q}}^-) &= \,
-\frac{{\B{1}{2}}^2}{\B{1}{3}}\ .
    \label{mhvqgqbar}
\eea
Here as always, all momenta $k_i$ are assumed to be complex which
ensures  that these 3-point amplitudes do not vanish on-shell
\cite{BCF:rec,BCFW:proof}.

In addition we need
to introduce two new primitive vertices for an off-shell vector boson coupled to a
$q\bar{q}$ pair. They are derived directly from the Feynman rules:
\def\sigmab{\bar{\sigma}}
\bea
    S_\mu(1_q^-,2_{\bar{q}}^+) &=& \,  \AA{1}{\sigma_\mu}{2}  \equiv
    \BB{2}{\sigmab_\mu}{1}\ , \nonumber \\
    S_\mu(1_q^+,2_{\bar{q}}^-) &=& \,  \BB{1}{\sigmab_\mu}{2} \equiv
    \AA{2}{\sigma_\mu}{1}\ .
    \label{fourfive}
\eea
Here, $\sigma_{\mu\,\alpha\dot{\alpha}}$ and
$\sigmab^{\dot{\alpha}\alpha}_{\mu}$ are the standard four Pauli matrices.
Our conventions for spinor contractions are summarised in the Appendix.

Action of parity symmetry reduces the number of independent currents.
Parity transformation is simply the complex conjugation
in terms of $S_{\alpha \dot\beta} := S_\mu \sigma^\mu_{\alpha \dot\beta}$,
\be
S_{\alpha\dot{\beta}}(1_q^{\lambda},2^{h_2},\ldots,(n-1)^{h_{n-1}},n_{\bar{q}}^{-\lambda}) =\,
    \left (
S_{\beta\dot{\alpha}}(1_q^{-\lambda},2^{-h_2},\ldots,(n-1)^{-h_{n-1}},n_{\bar{q}}^{\lambda})
    \right )^*.
    \label{eq:parity}
\ee
This formula is  generalised to multi-vector boson currents in an obvious way:
\be
\label{eq:paritygen}
S_{\alpha_1\dot{\beta}_1 \ldots \alpha_m\dot{\beta}_m } = \,
(S_{\beta_1\dot{\alpha}_1 \ldots \beta_m\dot{\alpha}_m })^*.
\ee

\subsection{Single Currents with $n=4,5,6$ partons}

In the four-parton case the recursion relation for $S_\mu$ reduces to,
\begin{equation}
    S_\mu(1_q^-,2^\pm,3_{\bar{q}}^+) =
    S_\mu(\widehat{1}_q^-,-\widehat{P}_{\bar{q}}^+)\frac{1}{s_{23}}
    \mathcal{A}(\widehat{P}_q^-,\widehat{2}^\pm,3_{\bar{q}}^+).
\end{equation}
We now have to choose which of the marked particles, $\widehat{1}$, $\widehat{2}$,
is $\hat{i}$ and which is $\hat{j}$.
In order to avoid boundary contributions according to \eqref{hconds} we must choose
$i=2$ for the $2^+$ helicity and $j=2$ for $2^-$ helicity. This results in,
\begin{align}
    S_\mu(1_q^-,2^+,3_{\bar{q}}^+) &=\,
    - \,\frac{\Aa{1}{\sigma_\mu P_V}{1}}{\A{1}{2}\A{2}{3}}\label{eq:4pt2}\, ,\\
    S_\mu(1_q^-,2^-,3_{\bar{q}}^+) &=\, \frac{\Bb{3}{P_V\sigma_\mu}{3}}{\B{1}{2}\B{2}{3}}\, ,
    \label{eq:4pt1}
\end{align}
where $P_V$ is the momentum of the vector boson.
Momentum conservation implies
$P_V=-p_1 -p_2-p_3.$
The two remaining helicity
configurations can be obtained by parity transformations.

\begin{figure}[t]
    \psfrag{1}{$1_q$}
    \psfrag{2}{$2$}
    \psfrag{3}{$3$}
    \psfrag{4}{$4_{\bar{q}}$}
    \psfrag{mu}{$\mu$}
    \begin{center}
        \includegraphics[width=12cm]{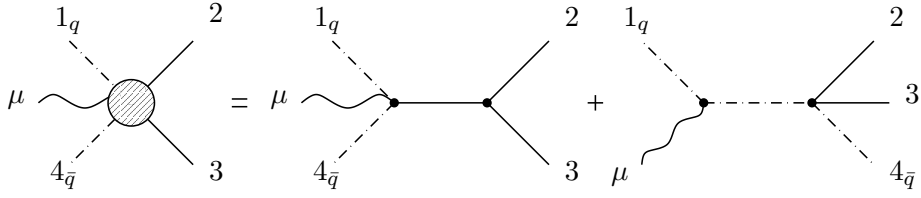}
    \end{center}
    \caption{Decomposition of the five-parton amplitude after applying the recursion relation.}
    \label{fig:vqqgg}
\end{figure}

Figure \ref{fig:vqqgg} shows the decomposition of the five-parton current. Just as in the four-parton
case here we mark the quark $\widehat{1}$ and adjacent gluon $\widehat{2}$,
\begin{align}
    S_\mu(1_q^-,2^+,3^+,4_{\bar{q}}^+) &=  
    \frac{
    \Aa{1}{\sigma_\mu P_V}{1}
    }{
    \A{1}{2}\A{2}{3}\A{3}{4}
    }\, , \label{eq:5ptmppp}\\
    S_{\mu}(1_q^-,2^-,3^+,4_{\bar{q}}^+) &=\,
    \frac{
            \B{1}{3}\Bb{3}{(1 + 2)P_V\sigmab_{\mu}(1 + 2)}{3}
    }
    {
            s_{123}\B{1}{2}\B{2}{3}\AA{4}{2 + 3}{1}
    }\nonumber \\
    &+\frac{
            \A{2}{4}\Aa{2}{(3 + 4)\sigmab_{\mu}P_V(3 + 4)}{2}
    }
    {
        s_{234}\A{2}{3}\A{3}{4}\AA{4}{2 + 3}{1}
    }\, ,\label{eq:5ptmmpp}\\
    S_{\mu}(1_q^-,2^+,3^-,4_{\bar{q}}^+) &=\,
    \frac{
            \B{2}{4}^3\Aa{1}{\sigma_{\mu}P_V}{1}
    }
    {
            s_{234}\B{2}{3}\B{3}{4}\AA{1}{2 + 3}{4}
    }\nonumber \\
    &+\frac{
            \A{1}{3}^3\Bb{4}{P_V \sigma_{\mu}}{4}
    }
    {
            s_{123}\A{1}{2}\A{2}{3}\AA{1}{2 + 3}{4}
    }\, .
    \label{eq:5ptmpmp}
\end{align}
The $S_\mu(+;++;-)$ configuration can be obtained from eq.~\eqref{eq:5ptmppp} by using the
``line reversal" symmetry,
\begin{equation}
    S_\mu(1_q^\lambda,2^{h_2},\ldots,n-1^{h_{n-1}},n_{\bar{q}}^{-\lambda}) =
    (-1)^{n+1}S_\mu(n_q^{-\lambda},n-1^{h_{n-1}},\ldots,2^{h_2},1_{\bar{q}}^\lambda) \, .\label{eq:LR}
\end{equation}
The other 4 helicity amplitudes can then by obtained via parity transformations.
Notice that if we
use both \eqref{eq:parity} and \eqref{eq:LR}
we can also relate \eqref{eq:4pt1} and \eqref{eq:4pt2}.

The 6-point amplitudes can be computed in much the same way.
We choose to mark massless particles in such a way as to generate the
most compact analytic expression.
The following amplitude was computed using $i=2$ and $j=1$:
\begin{align}
    S_{\mu}(1_q^-,2^+,3^+,4^+,5_{\bar{q}}^+) &=\, - \,
    \frac{
            \Aa{1}{\sigma_{\mu}P_V}{1}
    }
    {
            \A{1}{2}\A{2}{3}\A{3}{4}\A{4}{5}
    }\, .\label{eq:6ptmpppp}
\end{align}
The following expressions was derived using $i=3$ and $j=4$:
\begin{align}
    S_{\mu}(1_q^-,2^-,3^+,4^+,5_{\bar{q}}^+) &=
    -\frac{
            \A{2}{5}\Aa{2}{(3 + 4 + 5)P_V\sigma_{\mu}(3 + 4 + 5)}{2}
    }
    {
            s_{2345}\A{2}{3}\A{3}{4}\A{4}{5}\AA{5}{P_V}{1}
    }\nonumber \\
    &+\frac{
            \B{1}{3}\Bb{3}{(1 + 2)\sigma_{\mu}P_V(1 + 2)}{3}
    }
    {
            s_{123}\A{4}{5}\B{1}{2}\B{2}{3}\AA{4}{2 + 3}{1}
    }\nonumber \\
    &+\frac{
            \AA{2}{3 + 4}{1}\Aa{2}{(3 + 4)(1 + 2 + 3 + 4)\sigma_{\mu}P_V(1 + 2 + 3 + 4)(3 + 4)}{2}
    }
    {
            s_{1234}s_{234}\A{2}{3}\A{3}{4}\AA{5}{P_V}{1}\AA{4}{2 + 3}{1}
    }\, .\label{eq:6ptmmppp}
\end{align}
For the choice $i=2$ and $j=1$ we find:
\begin{align}
    S_{\mu}(1_q^-,2^+,3^-,4^+,5_{\bar{q}}^+) &=
    -\frac{
            \A{1}{3}^3\A{3}{5}\Aa{3}{(4 + 5)P_V\sigma_{\mu}(4 + 5)}{3}
    }
    {
            \A{1}{2}\A{2}{3}\A{3}{4}\A{4}{5}\Aa{5}{P_V(1 + 2)}{3}\Aa{3}{(4 + 5)P_V}{1}
    }\nonumber \\
    &+\frac{
            \A{3}{5}\AA{3}{4 + 5}{2}^3\Aa{1}{\sigma_{\mu}P_V}{1}
    }
    {
            s_{2345}s_{345}\A{3}{4}\A{4}{5}\Aa{3}{(4 + 5)P_V}{1}\AA{5}{3 + 4}{2}
    }\nonumber \\
    &-\frac{
            \A{1}{3}^3\AA{3}{1 + 2}{4}\Bb{4}{(1 + 2 + 3)\sigma_{\mu}P_V(1 + 2 + 3)}{4}
    }
    {
            s_{1234}s_{123}\A{1}{2}\A{2}{3}\Aa{5}{P_V(1 + 2)}{3}\AA{1}{2 + 3}{4}
    }\nonumber \\
    &-\frac{
            \B{2}{4}^4\Aa{1}{\sigma_{\mu}P_V}{1}
    }
    {
            s_{234}\AA{5}{3 + 4}{2}\AA{1}{2 + 3}{4}\B{2}{3}\B{3}{4}
    }\, .\label{eq:6ptmpmpp}
    \end{align}
Finally using $i=3$ and $j=2$ we find,
    \begin{align}
    S_{\mu}(1_q^-,2^+,3^+,4^-,5_{\bar{q}}^+) &=
    \frac{
            \A{1}{4}^3\Bb{5}{P_V\sigma_{\mu}}{5}
    }
    {
            s_{1234}\A{1}{2}\A{2}{3}\A{3}{4}\AA{1}{P_V}{5}
    }\nonumber \\
    &-\frac{
            \B{3}{5}^3\Aa{1}{\sigma_{\mu}P_V}{1}
    }
    {
            s_{345}\A{1}{2}\AA{2}{3 + 4}{5}\B{3}{4}\B{4}{5}
    }\nonumber \\
    &+\frac{
            \AA{4}{2 + 3}{5}^3\Aa{1}{\sigma_{\mu}P_V}{1}
    }
    {
            s_{234}s_{2345}\A{2}{3}\A{3}{4}\AA{2}{3 + 4}{5}\AA{1}{P_V}{5}
    }\, .\label{eq:6ptmppmp}
\end{align}
Using the parity and line reversal symmetries of eqs.~(\ref{eq:parity}) and (\ref{eq:LR})
we can easily obtain expressions 
for the other 12 helicity configurations.

All the amplitudes presented in this section have been numerically
checked against Feynman-diagram based calculations.

\subsection{$n$-point Currents}

\begin{figure}[t]
    \psfrag{p}{\small$+$}
    \psfrag{m}{\small$-$}
    \psfrag{1}{$\widehat{1}_q^-$}
    \psfrag{2}{$\widehat{2}^+$}
    \psfrag{3}{$3^+$}
    \psfrag{4}{$4^+$}
    \psfrag{nm1}{$n-1^+$}
    \psfrag{n}{$n_{\bar{q}}^+$}
    \psfrag{mu}{$\mu$}
    \begin{center}
        \includegraphics[width=10cm]{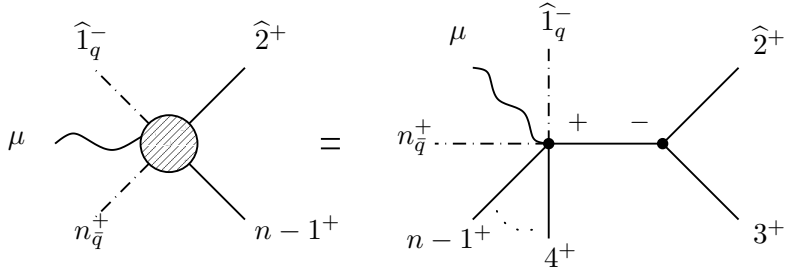}
    \end{center}
    \caption{Decomposition of the $n$-point vector boson current with $n-2$ positive helicity
    gluons.}
    \label{fig:allplusrec}
\end{figure}

It is also possible to construct single vector boson currents
with $n$ partons in the helicity configurations with maximal helicity
violation, next-to-maximal helicity violation and beyond.
The current for vector boson decaying to a quark
pair and any number of positive helicity gluons has been known
for some time~\cite{Berends:vector},
\def\denA#1{\prod_{\alpha=#1}^{n-1} \A{\alpha\,}{\alpha+1}}
\def\denB#1{\prod_{\alpha=#1}^{n-1} \B{\alpha\,}{\alpha+1}}
\begin{equation}
    S_\mu(1_q^-,2^+,\ldots,(n-1)^+,n_{\bar{q}}^+) = \, (-1)^n \,
    \frac{\Aa{1}{\sigma_\mu P_V}{1}}{\denA{1}}\, .
    \label{eq:allplus}
\end{equation}
As usual $P_V$ is the momentum of the vector boson and
$P_V=-p_1-\ldots-p_n$.
This can be easily proved by induction using on-shell recursion relations.
The fact that any pure QCD
amplitude with less than two negative helicities is zero guarantees that the
only contribution to the
$n$-point current involves an $(n-1)$-point current and an on-shell (complex)
3-gluon vertex, as shown in figure~\ref{fig:allplusrec}.
This is the first non-vanishing helicity amplitude and hence it is the MHV current.
For completeness, the other MHV-type currents are given by,
\bea
    S_\mu(1_q^+,2^+,\ldots,(n-1)^+,n_{\bar{q}}^-) &=& \, -(-1)^n \,
    \frac{\Aa{n}{\sigma_\mu P_V}{n}}{\denA{1}},\nonumber\\
    S_\mu(1_q^+,2^-,\ldots,(n-1)^-,n_{\bar{q}}^-) &=& \, \phantom{(-1)^n}-  
    \frac{\Bb{1}{ P_V \sigmab_\mu}{1}}{\denB{1}},\nonumber\\
    S_\mu(1_q^-,2^-,\ldots,(n-1)^-,n_{\bar{q}}^+) &=& \, \phantom{-(-1)^n} 
    \frac{\Bb{n}{P_V\sigmab_\mu}{n}}{\denB{1}}. 
\eea

\begin{figure}[t]
    \psfrag{p}{\small$+$}
    \psfrag{m}{\small$-$}
    \psfrag{1}{$\widehat{1}_q^-$}
    \psfrag{2}{$\widehat{2}^-$}
    \psfrag{j}{$j^+$}
    \psfrag{jp1}{$j+1^+$}
    \psfrag{nm1}{$n-1^+$}
    \psfrag{n}{$n_{\bar{q}}^+$}
    \psfrag{mu}{$\mu$}
    \psfrag{SUM}{\LARGE $\underset{j=3}{\overset{n-1}{\sum}}$}
    \begin{center}
        \includegraphics[width=14cm]{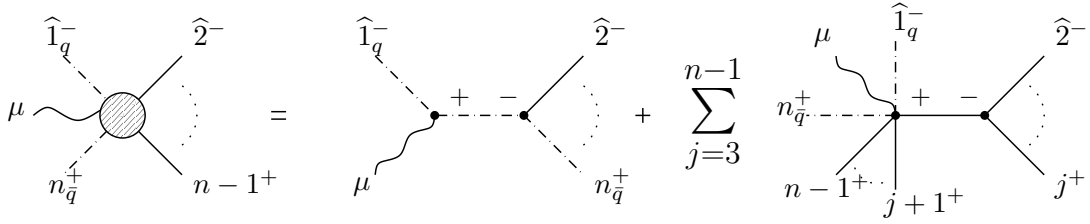}
    \end{center}
    \caption{Decomposition of the $n$-point vector boson current with $n-3$ positive helicity
    gluons and two adjacent negative helicities.}
    \label{fig:oneminusrec}
\end{figure}

It is interesting to note that 
eq.~(\ref{eq:allplus}) allows us to immediately write down compact expressions
for the NMHV currents with both adjacent and non-adjacent minuses. If we mark the two negative
helicity particles then each sub-amplitude in the recursion relation will contain at most 2 negative
helicities. Figure \ref{fig:oneminusrec} shows the decomposition into a sum of sub-diagrams.
We draw a pure QCD amplitude on the right of each diagram.
It must contain at least 2 negative helicity
particles and this fixes the helicity on the right of the
 propagator to be negative. Helicity conservation
then ensures that the vector current on the right has only one negative helicity, the marked
quark, and so is an MHV current. We can therefore use \eqref{eq:allplus} to write down the NMHV
current, marking $i=1$ and $j=2$:
\begin{align}
    S_\mu&(1_q^-,2^-,3^+,\ldots,(n-1)^+,n_{\bar{q}}^+)  = \frac{-(-1)^{n}}{\denA{2}} \Bigg (
    \frac{
    \Aa{2}{K_{3,n}P_V\sigma_\mu  K_{3,n}}{2}\A{2}{n}
    }{
    s_{2,n}\AA{n}{K_{2,n-1}}{1}
    } \nonumber \\
    &+ \sum_{j=3}^{n-1}\frac{
    \Aa{2}{K_{3,j}K_{1,j}P_V\sigmab_\mu K_{1,j}K_{3,j}}{2}\AA{2}{K_{3,j}}{1}\A{j\,}{j+1}
    }
    {
    s_{2,j}s_{1,j}\AA{j}{K_{2,j-1}}{1}\AA{j+1}{K_{2,j}}{1}
    }\Bigg ),
    \label{eq:NMHV1}
\end{align}
This expression is the $n$-parton
generalisation of equations \eqref{eq:5ptmmpp} and \eqref{eq:6ptmmppp}.

We can also consider the case where the helicity along the quark line is flipped. This is a special
case as we can still eliminate all contributions from NMHV vertices. The result is,
\begin{align}
    S_\mu&(1_q^-,2^+,\ldots,(n-2)^+,(n-1)^-,n_{\bar{q}}^+)  =
    \frac{(-1)^n}{\prod_{\alpha=1}^{n-2} \A{\alpha\,}{\alpha+1}} \Bigg (
    \frac{
    \A{1}{n-1}^3\Bb{n}{\sigmab_\mu P_V}{n}
    }{
    s_{1,n-1}\AA{1}{P_V}{n}
    }\nonumber\\
    &+\sum_{j=1}^{n-3}\frac{
    \AA{n-1}{K_{j+1,n-2}}{n}^3\Aa{1}{\sigma_\mu P_V}{1}\A{j\,}{j+1}
    }{
    s_{j+1,n}s_{j+1,n-1}\AA{j+1}{K_{j+1,n-1}}{n}\AA{j}{K_{j+1,n-1}}{n}
    }\Bigg ),
    \label{eq:NMHV1a}
\end{align}
matching equations \eqref{eq:5ptmpmp} and \eqref{eq:6ptmppmp} when $n=4$ and $5$ respectively.

\begin{figure}[t]
    \psfrag{p}{\small$+$}
    \psfrag{m}{\small$-$}
    \psfrag{1}{$\widehat{1}_q^-$}
    \psfrag{2}{$\widehat{2}^+$}
    \psfrag{3}{$3^-$}
    \psfrag{j}{$j^+$}
    \psfrag{jp1}{$j+1^+$}
    \psfrag{nm1}{$n-1^+$}
    \psfrag{n}{$n_{\bar{q}}^+$}
    \psfrag{mu}{$\mu$}
    \psfrag{SUM}{\LARGE $\underset{j=4}{\overset{n-1}{\sum}}$}
    \begin{center}
        \includegraphics[width=14cm]{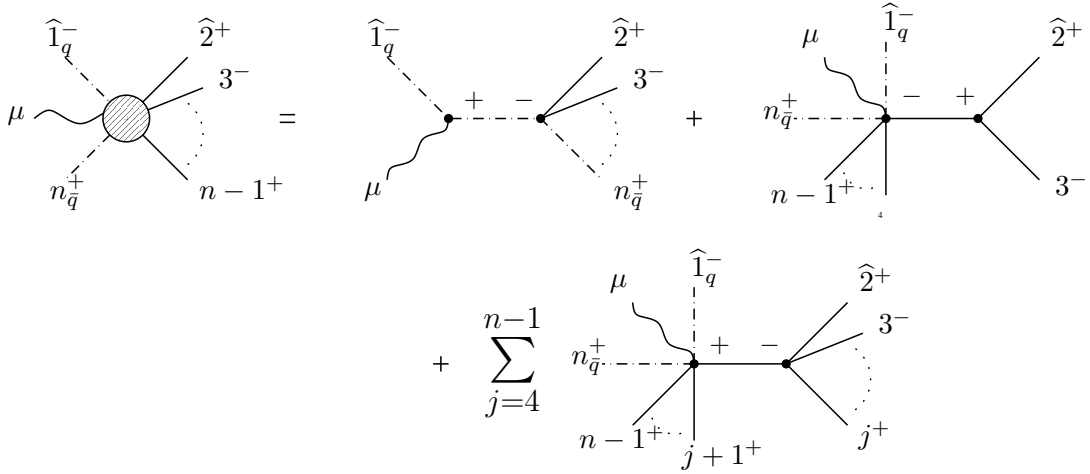}
    \end{center}
    \caption{Decomposition of the $n$-point vector boson current with $n-3$ positive helicity
    gluons and two negative helicities, separated by a single gluon with positive helicity. The
    second contribution involves the NMHV current with adjacent negative helicities given in
    (4.21). }
    \label{fig:oneminusrec2}
\end{figure}

For NMHV currents with non-adjacent negative helicities we can re-use the above result to find the
amplitude where the negative helicities are separated by one positive helicity. 
The corresponding diagrams are shown in Fig.~\ref{fig:oneminusrec2}
where we mark $i=2$
and $j=1$,
\begin{align}
    S_\mu&(1_q^-,2^+,3^-,4^+,\ldots,n-1^+,n_{\bar{q}}^+) =
    \frac{
    \Aa{1}{P_V\sigmab_\mu}{1}\AA{3}{K_{4,n}}{2}^3\A{3}{n}
    }{
    s_{2,n}s_{3,n}\Aa{1}{K_{2,n}K_{4,n}}{3}\AA{n}{K_{3,n-1}}{2}\denA{3}
    }\nonumber\\
    &+\sum_{j=4}^{n-1}\frac{
    \Aa{1}{\sigma_\mu P_V}{1}\AA{3}{K_{4,j}}{2}^4\A{j\,}{j+1}
    }{
    s_{2,j}s_{3,j}\Aa{1}{K_{2,j}K_{4,j}}{3}\AA{j}{K_{3,j-1}}{2}\AA{j+1}{K_{3,j}}{2}\denA{3}
    }\nonumber\\
    &+S_\mu(\wh{1}_q^-,\wh{P}^-,4^+,\ldots,n-1^+,n_{\bar{q}}^+)
    \frac{\A{3}{1}^3}{\A{1}{2}\A{2}{3}\bar{w}^2}
\end{align}
Substituting eq.~\eqref{eq:NMHV1} and simplifying the shifts results in the following expression:
\begin{align}
    S_\mu&(1_q^-,2^+,3^-,4^+,\ldots,n-1^+,n_{\bar{q}}^+) = \frac{1}{\denA{3}} \Bigg (
    \frac{
    \Aa{1}{P_V\sigmab_\mu}{1}\AA{3}{K_{4,n}}{2}^3\A{3}{n}
    }{
    s_{2,n}s_{3,n}\Aa{1}{K_{2,n}K_{4,n}}{3}\AA{n}{K_{3,n-1}}{2}
    }\nonumber\\
    &+\sum_{j=4}^{n-1}\frac{
    \Aa{1}{\sigma_\mu P_V}{1}\AA{3}{K_{4,j}}{2}^4\A{j\,}{j+1}
    }{
    s_{2,j}s_{3,j}\Aa{1}{K_{2,j}K_{4,j}}{3}\AA{j}{K_{3,j-1}}{2}\AA{j+1}{K_{3,j}}{2}
    }\nonumber\\
    &+\frac{
    \A{1}{3}^3\A{n}{3}\Aa{3}{K_{4,n}P_V\sigma_\mu K_{4,n}}{3}
    }{
    \Aa{1}{K_{2,n}K_{4,n}}{3}\Aa{n}{K_{1,n-1}K_{12}}{3}\A{1}{2}\A{2}{3}
    } \nonumber\\
    &+\sum_{j=4}^{n-1}\frac{
    \A{1}{3}^3\Aa{3}{K_{4,j}K_{1,j}P_V\sigmab_\mu K_{1,j} K_{4,j}}{3}
    \Aa{3}{K_{12}K_{4,j}}{3}\A{j\,}{j+1}
    }{
    s_{1,j}\Aa{1}{K_{2,j}K_{4,j}}{3}\Aa{j}{K_{1,j-1}K_{12}}{3}
    \Aa{j+1}{K_{1,j}K_{12}}{3}\A{1}{2}\A{2}{3}
    }\Bigg ).
    \label{eq:NMHV2}
\end{align}
Remaining NMHV currents can be constructed in a similar way.
We can keep adding an extra positive helicity separating the
negative helicities.
We have checked that the $n$-parton result in eq.~\eqref{eq:NMHV2}
agrees with eqs.~\eqref{eq:5ptmpmp} and
\eqref{eq:6ptmpmpp} for $n=4$ and $n=5$ respectively.

As a final example we consider the NNMHV current with three adjacent negative helicities.

\begin{figure}[t]
	\psfrag{mu}{$P_{V_1}$}
	\psfrag{1}{$1_q^-$}
	\psfrag{2}{$\wh{2}^-$}
	\psfrag{3}{$\wh{3}^-$}
	\psfrag{j}{$j^+$}
	\psfrag{jp1}{$j+1^+$}
	\psfrag{nm1}{$n-1^+$}
	\psfrag{n}{$n_{\bar{q}}^+$}
	\psfrag{m}{$\small -$}
	\psfrag{p}{$\small +$}
	\psfrag{SUM}{\LARGE $\underset{j=4}{\overset{n-1}{\sum}}$}
	\begin{center}
		\includegraphics[width=14cm]{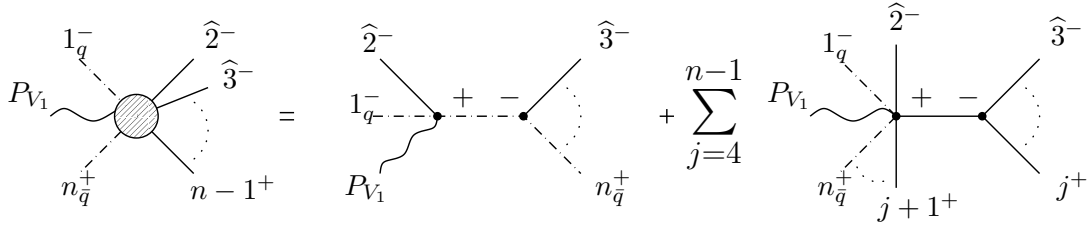}
	\end{center}
	\caption{Decomposition of the single vector boson current with three consecutive negative helicites. 
	The recursion re-uses the current with two consecutive nehative helicites.}
	\label{fig:twominus}
\end{figure}

By marking particles $i=2$ and $j=3$ we ensure that only the NMHV current \eqref{eq:NMHV1} is needed.
Explicit calculation yields,
\begin{align}
    S_\mu&(1_q^-,2^-,3^-,4^+,\ldots,n-1^+,n_{\bar{q}}^+) = \frac{(-1)^n}{\denA{3}}\Bigg (
    \frac{
    \A{3}{n}\Aa{3}{K_{4,n}P_V\sigma_\mu K_{4,n}}{3}
    }{
    s_{3,n}\B{1}{2}\AA{n}{K_{3,n-1}}{2}
    } \nonumber\\
    &+\sum_{j=4}^{n-1}\frac{
    \AA{3}{K_{4,j}}{1}\Aa{3}{K_{4,j}K_{1,j}P_V\sigma_\mu K_{1,j} K_{4,j}}{3}\A{j\,}{j+1}
    }{
    s_{3,j}s_{1,j}\B{1}{2}\AA{j+1}{K_{2,j}}{1}\AA{j}{K_{3,j-1}}{2}
    } \nonumber\\
    &-\sum_{j=4}^{n-1}\sum_{l=j+1}^{n-1}\frac{
    \AA{3}{K_{4,j}K_{2,j}K_{2,l}}{1}\Aa{3}{K_{4,j}K_{2,j}K_{2,l}K_{1,l}P_V\sigmab_\mu
    K_{1,l}K_{2,l}K_{2,j}K_{4,j}}{3}\A{j\,}{j+1}\A{l\,}{l+1}
    }{
    s_{1,l}s_{2,l}s_{2,j}s_{3,j}
    \AA{l}{K_{2,l-1}}{1}\AA{l+1}{K_{2,l}}{1}\AA{j+1}{K_{3,j}}{2}\AA{j}{K_{3,j-1}}{2}
    }\nonumber\\
    &-\sum_{j=4}^{n-1}\frac{
    \Aa{3}{K_{4,j}K_{2,j}}{n}\Aa{3}{K_{4,j}K_{2,j}K_{2,n}P_V\sigmab_\mu
    K_{2,n}K_{2,j}K_{4,j}}{3}\A{j\,}{j+1}
    }{
    s_{3,j}s_{2,n}s_{2,j}\AA{n}{K_{2,n-1}}{1}\AA{j+1}{K_{3,j}}{2}\AA{j}{K_{3,j-1}}{2}
    }
    \Bigg ).
    \label{eq:NNMHV}
\end{align}
We have explicitly checked eq.~\eqref{eq:NNMHV} for up to six partons.

By repeated use of the recursion formulae, further $n$-point currents may be obtained.

%%%%%%%%%%%%%%%%%%%%%%%%%%%%%%%%%%%%%%%%%%%%%%%%%

\section{Double Vector Boson Currents \label{sec:double}}

We now turn to double vector boson currents $S_{\mu\nu}$.
We start by considering the smallest amplitude of this type, the one with only two partons,
$S_{\mu\nu}(q,\bar{q})$.
\begin{figure}[t]
    \psfrag{q}{$2_{\bar{q}}$}
    \psfrag{qb}{$1_q$}
    \psfrag{w1}{$P_{V_1}$}
    \psfrag{w2}{$P_{V_2}$}
    \begin{center}
        \includegraphics[width=10cm]{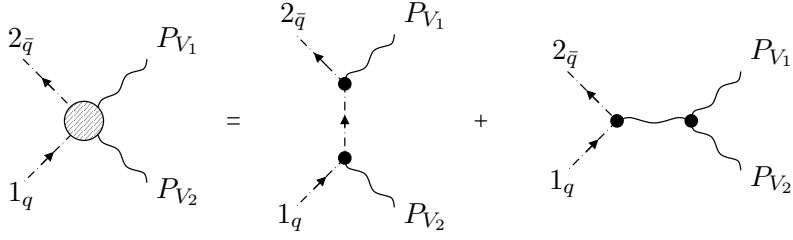}
    \end{center}
    \caption{The two Feynman diagrams contributing to the amplitude with two massive vector bosons and a
    quark pair.}
    \label{fig:VVqq1}
\end{figure}
One might expect that on-shell recursion relations can be used
to derive $S_{\mu\nu}(q,\bar{q})$
from two single vector boson amplitudes $S_{\mu}(q,\bar{q}).$
However there is a difficulty in writing down such a recursion relation.
We cannot mark the two massless particles in $S_{\mu\nu}(q,\bar{q})$ since
it is known that
marking adjacent massless quarks results in a
non-vanishing boundary contribution \cite{Luo:cmpct6pt} to the amplitude.
Choosing to mark massive particles also leads to (unnecessary)
technical
complications. It is actually much simpler to derive
the four-point amplitude $S_{\mu\nu}(q,\bar{q})$ from Feynman diagrams and
 use this four-point amplitude as a new primitive vertex in
further recursive calculations of
$S_{\mu\nu}(1_q,\ldots , n_{\bar{q}}).$

In fact, we will use a more elegant approach.
In general, there are two Feynman diagrams
contributing to $S_{\mu\nu}(q,\bar{q})$, as shown in figure \ref{fig:VVqq1}.
We could evaluate both diagrams and use the whole amplitude as a building block
for larger amplitudes, however, it is much more efficient to split the calculation into two parts in
order to re-use the single vector boson currents computed in section {\bf \ref{sec:single}}.

The first part corresponds to the second diagram in figure \ref{fig:VVqq1},
it contains the non-Abelian three-vertex of massive vector bosons.
We can compute such contributions to a generic
$ S_{\mu\nu}(1_q,\ldots,n_{\bar{q}})$
by contracting two single vector boson currents $S_{\mu}(1_q,\ldots,n'_{\bar{q}})$
of the previous section with the
colour ordered Feynman three-point vertex. This approach was used to calculate the non-Abelian
contribution to $S_{\mu\nu}(q,g,\bar{q})$ in reference \cite{Bern:EWcurrents}. Note that if one is dealing
with uncharged gauge bosons which have no self-coupling, for example Z bosons, this term is trivially zero.

The second part does not contain a non-Abelian three-vertex of vector bosons,
it corresponds to the first diagram in figure \ref{fig:VVqq1}.
This second Abelian contribution to a generic $S_{\mu\nu}(1_q,\ldots,n_{\bar{q}})$
can then be evaluated using
on-shell recursion relations.

Thus, guided by figure \ref{fig:VVqq1} we represent the
 colour ordered double current with $n$ partons in the form:
\begin{align}
    S_{\mu\nu} (1_q,\dots,n_{\bar{q}})&= \,
    T^{(3)}_{\mu\nu\rho}(P_{V_1},P_{V_2},-P)\, \frac{1}{(P^2-M_P^2)}\, S^{\rho}(1_q,\dots,n_{\bar{q}})
    \nonumber\\
    &+S_{\mu\nu}^{Abelian}(1_q,\dots,n_{\bar{q}}).
    \label{eq:VVdecomp}
\end{align}
Here
\begin{equation}
    T^{(3)}_{\mu_1\mu_2\mu_3}(p_1,p_2,p_3) =
    g_{\mu_1\mu_2}(p_1-p_2)_{\mu_3}+g_{\mu_2\mu_3}(p_2-p_3)_{\mu_1}+g_{\mu_3\mu_1}(p_3-p_1)_{\mu_2}
\end{equation}
is the colour-ordered three-vertex of massive vector bosons,
with all momenta defined to be in-going.

The {\it primitive vertex} for the Abelian contribution is given by the first Feynman diagram in figure
\ref{fig:VVqq1} which evaluates to:
\begin{align}
    S^{Abelian}_{\mu\nu}(1_q^-,2_{\bar{q}}^+) &=
    \frac{1}{s_{1P_{V_2}}}\AA{1}{\sigma_{\nu}(1+P_{V_2})\sigma_{\mu}}{2}.
    \label{eq:VVqq}
\end{align}
%Using the formulation of recursion relations for massive particles \cite{mrscalar} it should be
%possible to reproduce this four point amplitude by marking the massive particles. Marking the
%adjacent quark and boson results in a zero contribution as there is no vector boson self-coupling in
%the Abelian contribution. Marking the particles opposite each other results in a single diagram
%which, un-shifted, would exactly match the Feynman
%graph. However, following the prescription of \cite{mrscalar}, it turns out there is a non-zero
%boundary contribution. This can be seen explicitly by looking at the behaviour of the amplitude
%\eqref{eq:VVqq} in the $z \to \infty$ limit after
%continuation to the complex plane via $P_{V_1}\to P_{V_1}+z|1\ra \la
%1|P_{V_1}$ and $p_1 \to p_1 - z|1\ra \la 1|P_{V_1}$. So it appears that it is not possible to form a
%simple recursive calculation for this 4-point amplitude. However evaluation of the single Feynman diagram
%allows us to compute higher point currents using this amplitude as a primitive vertex.

The remaining non-Abelian part of this four-point amplitude is determined by the first
line of \eqref{eq:VVdecomp}
\be
S_{\mu\nu}^{non-Abelian} (1_q,2_{\bar{q}})= \,
    T^{(3)}_{\mu\nu\rho}(P_{V_1},P_{V_2},-P)\, \frac{1}{(P^2-M_P^2)}\,
S^{\rho}(1_q,2_{\bar{q}})
\ee
 where $S^{\rho}(1_q,2_{\bar{q}})$ is given in \eqref{fourfive}.

In general one needs to determine only the Abelian components
of the $n$-parton double currents $S_{\mu\nu} (1_q,\dots,n_{\bar{q}})$,
the non-Abelian components are fully determined by
the first
line of \eqref{eq:VVdecomp} in terms of the known single currents.

Abelian components are characterised by having massive vector bosons only
on external lines, and they can always be calculated recursively.
In general, in order to calculate the Abelian part of any
double current, $S^{Abelian}_{\mu\nu}(1_q,2,\ldots,n_{\bar{q}}^+),$
one needs to draw all recursive decompositions of this current
such that the internal line is a quark or a gluon and not a massive vector boson.

\begin{figure}[t]
    \psfrag{mu}{$P_{V_1}$}
    \psfrag{nu}{$P_{V_2}$}
    \psfrag{1}{$\widehat{1}_q$}
    \psfrag{2}{$\widehat{2}$}
    \psfrag{3}{$3_{\bar{q}}$}
    \begin{center}
        \includegraphics[width=15cm]{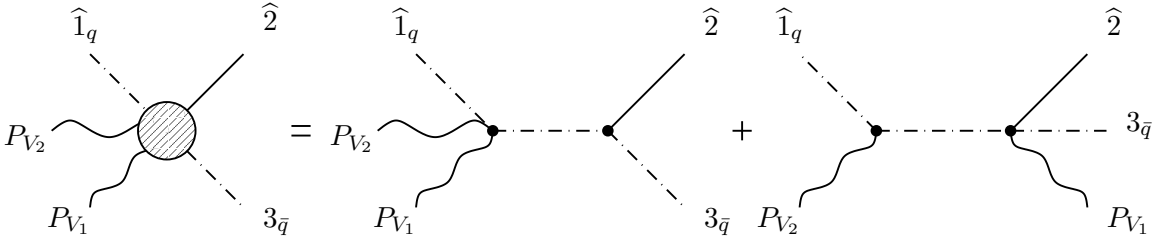}
    \end{center}
    \caption{Recursive decomposition of the five-point Abelian amplitude for two vector bosons,
    a quark pair and a gluon
    using the recursion relation.}
    \label{fig:5ptvvqqg}
\end{figure}
First we calculate the two five-point amplitudes by marking the quark and adjacent gluon. The
recursion relation for $S^{Abelian}_{\mu\nu}(1_q,2,3_{\bar{q}})$ is depicted
in figure \ref{fig:5ptvvqqg}. It yields,
\begin{align}
    S^{Abelian}_{\mu\nu}(1_q^-,2^-,3_{\bar{q}}^+) &=
    +\frac{
    \B{1}{3}\Bb{3}{\sigmab_{\mu}(1 + P_{V_2})\sigmab_{\nu}(1 + 2)}{3}
    }
    {
            \B{1}{2}\B{2}{3}\Bb{3}{P_{V_1}P_{V_2}}{1}
    }\nonumber \\
    &+\frac{
            \AA{2}{\sigma_{\mu}}{3}\Bb{1}{\sigmab_{\nu}(1 + 2)}{3}
    }
    {
            \B{1}{2}\Bb{3}{P_{V_1}P_{V_2}}{1}
    }\nonumber \\
    &-\frac{
            \Aa{2}{(1 + P_{V_2})\sigmab_{\nu}P_{V_2}(1 + P_{V_2})}{2}\Bb{3}{P_{V_1}\sigma_{\mu}}{3}
    }
    {
            s_{1P_{V_2}}s_{3P_{V_1}}\Bb{3}{P_{V_1}P_{V_2}}{1}
    }\,,\\
    S^{Abelian}_{\mu\nu}(1_q^-,2^+,3_{\bar{q}}^+) &=
    \frac{
            \A{1}{3}\Aa{1}{(2 + 3)\sigmab_{\mu}(1 + P_{V_2})\sigmab_{\nu}}{1}
    }
    {
            \A{1}{2}\A{2}{3}\Aa{1}{P_{V_2}P_{V_1}}{3}
    }\nonumber \\
    &+\frac{
            \AA{1}{\sigma_{\nu}}{2}\Aa{1}{(2 + 3)\sigmab_{\mu}}{1}
    }
    {
            \A{1}{2}\Aa{1}{P_{V_2}P_{V_1}}{3}
    }\nonumber \\
    &-\frac{
            \Aa{1}{P_{V_2}\sigmab_{\nu}}{1}\Bb{2}{(1 + P_{V_2})P_{V_1}\sigmab_{\mu}(1 + P_{V_2})}{2}
    }
    {
        s_{1P_{V_2}}s_{3P_{V_1}}\Aa{1}{P_{V_2}P_{V_1}}{3}
    }\,.
\end{align}
The $S_{\mu\nu}(+;\pm;-)$ configurations can be obtained from 
eqs.~\eqref{eq:5ptmppp} by using either parity \eqref{eq:paritygen} or 
the ``line reversal" symmetry,
\begin{equation}
    S_{\mu\nu}(1_q^\lambda,2^{h_2},\ldots,n-1^{h_{n-1}},n_{\bar{q}}^{-\lambda}) =
    (-1)^{n+1}S_{\nu\mu}(n_q^{-\lambda},n-1^{h_{n-1}},\ldots,2^{h_2},1_{\bar{q}}^\lambda) \, .\label{eq:LRgen}
\end{equation}

Finally we give results for the six point amplitudes
$S^{Abelian}_{\mu\nu}(1_q,2,3,4_{\bar{q}}).$
Taking the generalised parity relation \eqref{eq:paritygen} and the line-reversal identity \eqref{eq:LRgen}, 
there are three independent helicity configurations.
Again we use on-shell recursion relations and mark
the quark and adjacent gluon. Choosing $i=2$ and $j=1$ we find,
\begin{align}
    S^{Abelian}_{\mu\nu}&(1_q^-,2^+,3^+,4_{\bar{q}}^+) =
    -\frac{
            \A{1}{4}\AA{1}{\sigma_{\nu}}{2}\Aa{1}{(2 + 3 + 4)\sigmab_{\mu}}{1}
    }
    {
            \A{1}{2}\A{1}{3}\A{3}{4}\Aa{1}{P_{V_2}P_{V_1}}{4}
    }\nonumber \\
    &-\frac{
            \A{1}{4}\Aa{1}{(2 + 3 + 4)\sigmab_{\mu}(1 + P_{V_2})\sigmab_{\nu}}{1}
    }
    {
            \A{1}{2}\A{2}{3}\A{3}{4}\Aa{1}{P_{V_2}P_{V_1}}{4}
    }\nonumber \\
    &+\frac{
            \Aa{1}{(2 + 3)\sigmab_{\nu}}{1}\Aa{1}{(P_{V_1} + P_{V_2})\sigmab_{\mu}}{1}
    }
    {
            \A{1}{2}\A{1}{3}\A{2}{3}\Aa{1}{P_{V_2}P_{V_1}}{4}
    }\nonumber \\
    &+\frac{
            \Aa{1}{P_{V_2}\sigmab_{\nu}}{1}\Aa{1}{(2 + 3)(4 + P_{V_1})P_{V_1}\sigmab_{\mu}(4 + P_{V_1})(2 + 3)}{1}
    }
    {
            s_{4P_{V_1}}\A{1}{2}\A{2}{3}\Aa{1}{P_{V_2}P_{V_1}}{4}\Aa{1}{P_{V_2}(4 + P_{V_1})}{3}
    }\nonumber \\
    &+\frac{
            \Aa{1}{P_{V_2}\sigmab_{\nu}}{1}\Bb{2}{(1 + P_{V_2})P_{V_1}\sigmab_{\mu}(1 + P_{V_2})}{2}
    }
    {
            s_{1P_{V_2}}s_{12P_{V_2}}\A{3}{4}\Aa{1}{P_{V_2}(4 + P_{V_1})}{3}
    },
 \end{align}
 and
\begin{align}
    S^{Abelian}_{\mu\nu}&(1_q^-,2^+,3^-,4_{\bar{q}}^+) =
    -\frac{
            \A{1}{3}^3\AA{3}{1 + 2}{4}\Bb{4}{\sigmab_{\mu}(1 + P_{V_2})\sigmab_{\nu}(1 + 2 + 3)}{4}
    }
    {
            s_{123}\A{1}{2}\A{2}{3}\AA{1}{2 + 3}{4}\AA{3}{(1 + 2)P_{V_2}P_{V_1}}{4}
    }\nonumber \\
    &-\frac{
            \A{1}{3}^3\Aa{3}{(4 + P_{V_1})\sigmab_{\nu}P_{V_2}(4 + P_{V_1})}{3}\Bb{4}{P_{V_1}\sigma_{\mu}}{4}
    }
    {
            s_{4P_{V_1}}\A{1}{2}\A{2}{3}\Aa{1}{P_{V_2}(4 + P_{V_1})}{3}\AA{3}{(1 + 2)P_{V_2}P_{V_1}}{4}
    }\nonumber \\
    &-\frac{
            \A{1}{3}^2\AA{1}{\sigma_{\mu}}{4}\AA{3}{1 + 2}{4}\Bb{2}{\sigmab_{\nu}(1 + 2 + 3)}{4}
    }
    {
            s_{123}\A{1}{2}\AA{1}{2 + 3}{4}\AA{3}{(1 + 2)P_{V_2}P_{V_1}}{4}
    }\nonumber \\
    &+\frac{
            \A{1}{3}^2\AA{3}{\sigma_{\mu}}{4}\AA{3}{(1 + 2)\sigmab_{\nu}(1 + 2 + 3)}{4}
    }
    {
            s_{123}\A{1}{2}\A{2}{3}\AA{3}{(1 + 2)P_{V_2}P_{V_1}}{4}
    }\nonumber \\
    &-\frac{
            \B{2}{4}^3\AA{1}{3 + 4}{2}\Aa{1}{(2 + 3 + 4)\sigmab_{\mu}(1 + P_{V_2})\sigmab_{\nu}}{1}
    }
    {
            s_{234}\AA{1}{P_{V_2}P_{V_1}(3 + 4)}{2}\AA{1}{2 + 3}{4}\B{2}{3}\B{3}{4}
    }\nonumber \\
    &-\frac{
            \B{2}{4}^3\AA{1}{\sigma_{\nu}}{2}\Aa{1}{(2 + 3 + 4)\sigmab_{\mu}}{1}
    }
    {
            \AA{1}{P_{V_2}P_{V_1}(3 + 4)}{2}\AA{1}{2 + 3}{4}\B{2}{3}\B{3}{4}
    }\nonumber \\
    &-\frac{
            \B{2}{4}^3\Aa{1}{P_{V_2}\sigmab_{\nu}}{1}\Bb{2}{(1 + P_{V_2})P_{V_1}\sigmab_{\mu}(1 + P_{V_2})}{2}
    }
    {
            s_{1P_{V_2}}\AA{1}{P_{V_2}P_{V_1}(3 + 4)}{2}\B{2}{3}\B{3}{4}\Bb{2}{(1 + P_{V_2})P_{V_1}}{4}
    }\nonumber \\
    &-\frac{
            \AA{3}{1 + P_{V_2}}{2}^3\Aa{1}{P_{V_2}\sigmab_{\nu}}{1}\Bb{4}{P_{V_1}\sigma_{\mu}}{4}
    }
    {
        s_{4P_{V_1}}s_{1P_{V_2}}s_{12P_{V_2}}\Aa{1}{P_{V_2}(4 + P_{V_1})}{3}\Bb{2}{(1 + P_{V_2})P_{V_1}}{4}
    }.
\end{align}
For the last amplitude we choose $i=1$ and $j=2$ which yields  
the following expression,
\begin{align}
    S&^{Abelian}_{\mu\nu}(1_q^-,2^-,3^+,4_{\bar{q}}^+) =
    -\frac{
            \A{2}{4}\BB{1}{\sigmab_{\nu}(P_{V_1} + P_{V_2})(3 + 4)}{2}\Aa{2}{(3 + 4)\sigmab_{\mu}}{2}
    }
    {
            \A{2}{3}\A{3}{4}\AA{2}{(3 + 4)P_{V_1}P_{V_2}}{1}\AA{4}{2 + 3}{1}
    }\nonumber \\
    &-\frac{
            \A{2}{4}\AA{2}{3 + 4}{1}\Aa{2}{(3 + 4)\sigmab_{\mu}(1 + P_{V_2})\sigmab_{\nu}(P_{V_1} + P_{V_2})(3 + 4)}{2}
    }
    {
        s_{234}\A{2}{3}\A{3}{4}\AA{2}{(3 + 4)P_{V_1}P_{V_2}}{1}\AA{4}{2 + 3}{1}
    }\nonumber \\
    &+\frac{
            \A{2}{4}\Aa{2}{(1 + P_{V_2})\sigmab_{\nu}P_{V_2}(1 + P_{V_2})}{2}\Aa{2}{(3 + 4)P_{V_1}\sigma_{\mu}(3 + 4)}{2}
    }
    {
            s_{1P_{V_2}}\A{2}{3}\A{3}{4}\Aa{2}{(1 + P_{V_2})P_{V_1}}{4}\AA{2}{(3 + 4)P_{V_1}P_{V_2}}{1}
    }\nonumber \\
    &-\frac{
            \B{1}{3}\BB{3}{1+2}{4}\Bb{3}{(1 + 2)(P_{V_1} + P_{V_2})\sigmab_{\mu}(1 + P_{V_2})\sigmab_{\nu}(1 + 2)}{3}
    }
    {
            s_{123}\AA{4}{2 + 3}{1}\B{1}{2}\B{2}{3}\BB{3}{(1 + 2)P_{V_2}P_{V_1}}{4}
    }\nonumber \\
    &-\frac{
            \B{1}{3}\Bb{3}{(1 + 2)P_{V_2}\sigmab_{\nu}(1 + 2)}{3}\Bb{3}{(4 + P_{V_1})P_{V_1}\sigmab_{\mu}(4 + P_{V_1})}{3}
    }
    {
        s_{4P_{V_1}}\B{1}{2}\B{2}{3}\Bb{3}{(4 + P_{V_1})P_{V_2}}{1}\BB{3}{(1 + 2)P_{V_2}P_{V_1}}{4}
    }\nonumber \\
    &-\frac{
            \AA{2}{1 + P_{V_2}}{3}\Aa{2}{(1 + P_{V_2})\sigmab_{\nu}P_{V_2}(1 + P_{V_2})}{2}\Bb{3}{(4 + P_{V_1})P_{V_1}\sigmab_{\mu}(4 + P_{V_1})}{3}
    }
    {
        s_{4P_{V_1}}s_{1P_{V_2}}s_{12P_{V_2}}\Aa{2}{(1 + P_{V_2})P_{V_1}}{4}\Bb{1}{P_{V_2}(4 + P_{V_1})}{3}
    }\nonumber \\
    &-\frac{
            \BB{3}{(1 + 2)(P_{V_1} + P_{V_2})\sigmab_{\mu}}{2}\BB{3}{1+2}{4}\Bb{1}{\sigmab_{\nu}(1 + 2)}{3}
    }
    {
            s_{123}\AA{4}{2 + 3}{1}\B{1}{2}\BB{3}{(1 + 2)P_{V_2}P_{V_1}}{4}
    }\nonumber \\
    &+\frac{
            \Bb{3}{\sigmab_{\nu}(1 + 2)}{3}\Bb{3}{(1 + 2)(P_{V_1} + P_{V_2})\sigmab_{\mu}(1 + 2)}{3}
    }
    {
            s_{123}\B{1}{2}\B{2}{3}\BB{3}{(1 + 2)P_{V_2}P_{V_1}}{4}
    }.
\end{align}

The procedure described here can straightforwardly be generalised to processes
involving three or more vector bosons.   In each case, there will be a mixture of 
terms that either involve a triple or quartic gauge boson vertex (non-Abelian) or
a new multi-gauge boson current (Abelian).  The non-abelian  contribution is
straightfoward and involves currents with coupling  that couples currents
involving fewer gauge bosons.  These are in principle known.    For each additional vector boson, the Abelian
contribution  must be recomputed.   There will be a new primitive vertex which can
be obtained directly from the single (colour-ordered) Feynman diagram.

\begin{figure}[t]
	\psfrag{1}{$1_q$}
	\psfrag{2}{$2_{\bar{q}}$}
	\psfrag{v1}{$P_{V_1}$}
	\psfrag{v2}{$P_{V_2}$}
	\psfrag{v3}{$P_{V_3}$}
	\begin{center}
		\includegraphics[width=14cm]{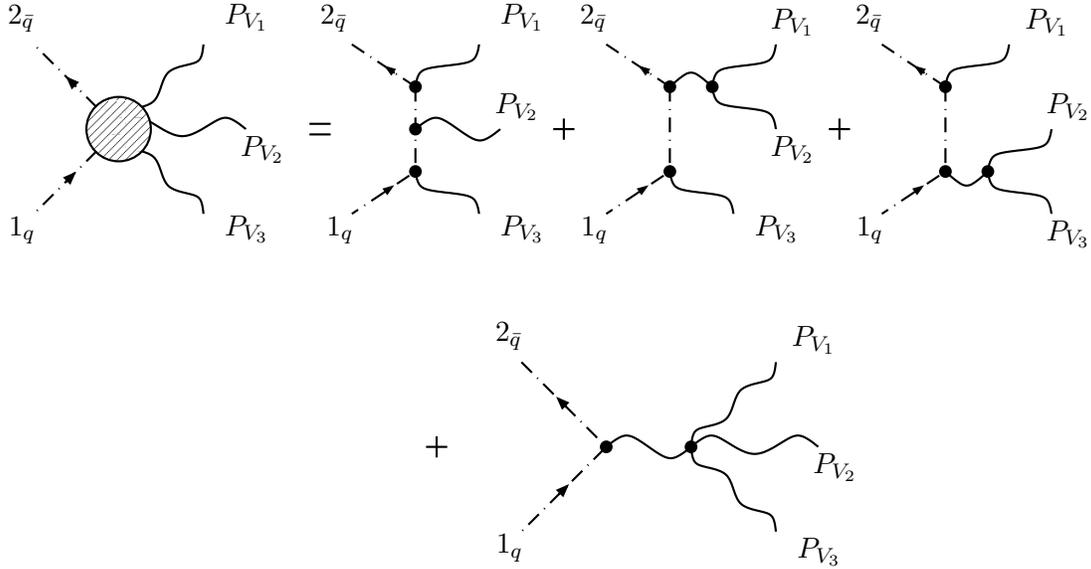}
	\end{center}
	\caption{Contributions to the amplitude with three massive vector bosons and a quark anti-quark
	pair.}
	\label{fig:VVVqq}
\end{figure}

This is illustrated in Fig.~\ref{fig:VVVqq} for three vector bosons. 
The first diagram yields a new primitive vertex $S_{\mu_1\mu_2\mu_3} (1_q,2_{\bar{q}})$
which forms the seed for recursively calculating the
Abelian contribution to the amplitude.   The other three (Non-Abelian)
graphs can be straightforwardly 
obtained by reusing the single and double vector boson currents.
The colour ordered triple current with $n$ partons is thus,
\begin{align}
    S_{\mu_1\mu_2\mu_3} (1_q,\dots,n_{\bar{q}})&= \,
    S_{\mu_1\mu_2\mu_3}^{Abelian}(1_q,\dots,n_{\bar{q}})    \nonumber\\
    &+T^{(3)}_{\mu_1\mu_2\rho}(P_{V_1},P_{V_2},-P_{12})\, 
    \frac{1}{(P_{12}^2-M_{P_{12}}^2)}\, S^{\rho\mu_3}(1_q,\dots,n_{\bar{q}})\nonumber\\
    &+T^{(3)}_{\mu_2\mu_3\rho}(P_{V_2},P_{V_3},-P_{23})\, 
    \frac{1}{(P_{23}^2-M_{P_{23}}^2)}\, S^{\mu_1\rho}(1_q,\dots,n_{\bar{q}})\nonumber\\
    &+T^{(4)}_{\mu_1\mu_2\mu_3\rho}(P_{V_1},P_{V_2},P_{V_3},-P_{123})\, 
    \frac{1}{(P_{123}^2-M_{P_{123}}^2)}\, S^{\rho}(1_q,\dots,n_{\bar{q}}).
    \label{eq:VVVdecomp}
\end{align}
The colour ordered quartic gauge boson vertex is given by,
\begin{equation}
    T^{(4)}_{\mu_1\mu_2\mu_3\mu_4}(p_1,p_2,p_3,p_4) =
     2g_{\mu_1\mu_3}g_{\mu_2\mu_4}-g_{\mu_1\mu_3}g_{\mu_2\mu_4}-g_{\mu_1\mu_2}g_{\mu_3\mu_4}.
\end{equation}

%%%%%%%%%%%%%%%%%%%%%%%%%%%%%%%%%%%%%%%%%%%%%%%%%%%%

\section{Recursion Relations for Massive Particles with Spin on Internal Lines \label{sec:spin}}

So far we have been considering application of recursion relations where
massive particles with spin were absent from the internal lines. In other words,
we have been able to set up the recursive calculations of double vector boson currents
in such a way that the
massive vector bosons were playing the role of `external sources' in the left and in the right
hand vertices, but were not propagating through the recursive diagram.
We now would like to show how to use recursion relations also for propagating massive
particles. In our earlier work \cite{mrscalar} we have accomplished this for massive scalars,
and now we want to generalise this approach to massive particles with spin.

The main difference between internal massive scalars of Ref.~\cite{mrscalar}
and internal massive fermions of vector bosons is that the latter have more than
one polararisation or spin state.
In the standard
recursion relation \eqref{eq:rrel} all particles are assumed to be in a state with
fixed helicity, and there is a summation over all these states.
We want to avoid using helicity states for internal massive particles
and instead to use a more natural basis of states

In this section we will describe a new way to implement the recursion relation
in this case and will illustrate its use by
calculating a simple amplitude of two heavy quarks scattering into two gluons.

The main point here is that the sum over helicities $h$ of the internal particle in
the standard recursion relation,
\begin{align}
\mathcal{A}_n (p_1,\ldots,p_n) =\,
    \sum_{{\rm partitions}}\sum_{h}
        & \mathcal{A}_L(p_r,\ldots,\wh{p}_i,\ldots,p_s,-\wh{P}^h) \,
    \frac{1}{P^2-m_P^2} \nonumber \\
    &\times \mathcal{A}_R(\wh{P}^{-h},p_{s+1},\ldots,\wh{p}_j,\ldots,p_{r-1}) \ ,
\end{align}
can be replaced by the sum over all of the spin states, rather than
 helicity quantum numbers which are not well suited for massive particles.
So, we first replace the sum over helicities by the sum over appropriately defined spin states.
For massive fermions this is the conventional spin
sum:
\begin{align}
    \sum_{s=1,2} u_s(p)\bar{u}_s(p) &= \s{p} + m_p \\
    \sum_{s=1,2} v_s(p)\bar{v}_s(p) &= \s{p} - m_p
    \label{eq:spinsum}
\end{align}
The remaining spinors and polarisation vectors of the
external massive particles can be left unfixed and simplified after squaring the amplitude with the
spin sums in the conventional way.

Using this in the recursion relation in which a massive quark propagates between the two diagrams
we have
\begin{align}
    &\sum_{s}
    \mathcal{A}_L(p_{r;\bar{q}},\ldots,\wh{p}_i,\ldots,p_s,-\wh{P}^s_q)
    \frac{1}{P^2-m_P^2}
    \mathcal{A}_R(\wh{P}^{-s}_{\bar{q}},p_{s+1},\ldots,\wh{p}_j,\ldots,p_{r-1;q})
    \nonumber\\
    =&
    \mathcal{A}_L(p_{r;\bar{q}},\ldots,\wh{p}_i,\ldots,p_s,-\wh{P}^{*})
    \frac{\wh{\s{P}}+m_P}{P^2-m_P^2}
    \mathcal{A}_R(\wh{P}^{*},p_{s+1},\ldots,\wh{p}_j,\ldots,p_{r-1;q})
\end{align}
where $P^*$ indicates the external spinor wave-function
has been stripped off this amplitude. In this way we can
use the benefits of using the recursion relations to provide reasonably compact formulae for
amplitudes with massive particles.

%%%%%%%%%%%%%%%%%%%%%%%%%%%%%%%%%%%%%%

\subsection{Example: Calculation of $\mathcal{A}_4(1_t,2,3,4_{\bar{t}})$ \label{sec:ferms}}

\begin{figure}[t]
    \psfrag{t}{$1_t$}
    \psfrag{tb}{$4_{\bar{t}}$}
    \psfrag{3}{$3$}
    \psfrag{2}{$2$}
    \begin{center}
        \includegraphics[width=10cm]{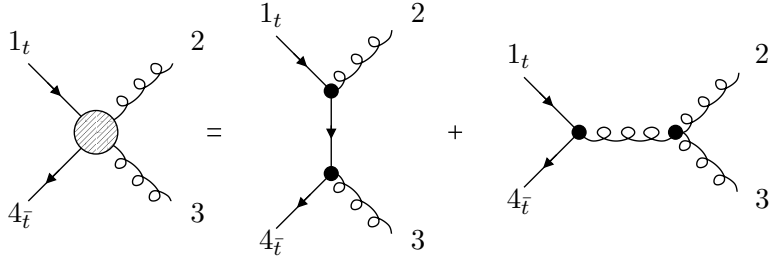}
    \end{center}
    \caption{The two Feynman diagrams contributing to the amplitude
    $\mathcal{A}_4(1_t,2,3,4_{\bar{t}})$.}
    \label{fig:ttgg}
\end{figure}

We now compute the four point amplitude of a top quark pair scattering to two gluons as an example
of the method described above.
We use on-shell recursion relations and mark two massless gluons.
This leads to a single recursive diagram with a massive fermion propagator.
We will show that the contribution of this single diagram precisely matches
the two Feynman diagrams for this process shown in figure \ref{fig:ttgg}. With all particles outgoing
the recursion relation result is,
\begin{equation}
    \mathcal{A}(1_t,2,3,4_{\bar{t}}) =\, \frac{1}{(P^2-m_t^2)}\, \bar{u}(p_1)\s{\e}(p_2,\xi_2)\left(\sum_s
    u_s(\wh{P})\bar{u}_s(\wh{P})\right)
    \s{\e}(p_3,\xi_3)v(p_4).
\end{equation}
Here $P=p_1+p_2$ is the momentum on the internal line and $\xi_2$, $\xi_3$ are reference spinors
necessary to specify gluon polarisation vectors $\epsilon^{\pm}$.
We will use the Weyl representation of the Dirac $\gamma$-matrices and polarisation
vectors,
\begin{equation}
    \gamma_\mu =
    \begin{pmatrix}
    0 & \sigma_\mu \\
    \bar{\sigma}_\mu & 0
    \end{pmatrix},
    \quad
    \s{\e}^+(p,\xi) = \frac{1}{\A{\xi}{p}}
    \begin{pmatrix}
    0 & |\xi\ra[p| \\
    |p]\la\xi| & 0
    \end{pmatrix},
    \quad
    \s{\e}^-(p,\xi) = \frac{1}{[\xi p]}
    \begin{pmatrix}
    0 & |p\ra[\xi|\\
    |\xi]\la p| & 0
    \end{pmatrix}.
\end{equation}
First we consider the case where the gluons have opposite helicity,
$\mathcal{A}(1_t,2^-,3^+,4_{\bar{t}})$. It is convenient to choose
$\xi_2=p_3$ and $\xi_3=p_2$ so that,
\begin{equation} \label{eqnsmth}
    \mathcal{A}(1_t,2^-,3^+,4_{\bar{t}})=\, \frac{1}{(P^2-m_t^2)s_{23}}\, \bar{u}(p_1)
    \begin{pmatrix}
        0 & |\wh{2}\ra [3| \\
        |3]\la\wh{2}| & 0
    \end{pmatrix}
    \begin{pmatrix}
        m_t & \s{\wh{P}} \\
        \s{\wh{P}} & m_t
    \end{pmatrix}
    \begin{pmatrix}
        0 & |2\ra [\wh{3}| \\
        |\wh{3}]\la2| & 0
    \end{pmatrix}
    v(p_4).
\end{equation}
We choose the marking prescription $i=3$ and $j=2$
and this ensures that the shifts on the polarisation vectors disappear.
It can also
be seen that the shift in $\widehat{P}$ is also killed by either of the two polarisation vectors and hence
we can erase all the hats in equation \eqref{eqnsmth}.
This is then exactly equivalent to the first diagram of figure \ref{fig:ttgg}.
It can be easily
shown using $\e^-(1,2)\cdot\e^+(2,1)=0$ that, with this particular choice of reference momenta,
that the remaining second
Feynman diagram gives a vanishing contribution
and so our recursion relation result is in agreement with the Feynman diagrams answer.

The amplitude with both gluons of negative helicity $\mathcal{A}(1_t,2^+,3^+,4_{\bar{t}})$
is of a non-MHV type and it provides another interesting test of the
recursion relation, which this time requires a little algebra.
The recursion relation reads:
\begin{equation}
    \mathcal{A}(1_t,2^+,3^+,4_{\bar{t}})=\frac{1}{(P^2-m_t^2)\A{2}{3}^2}\bar{u}(p_1)
    \begin{pmatrix}
        0 & |3\ra [\wh{2}| \\
        |\wh{2}]\la3| & 0
    \end{pmatrix}
    \begin{pmatrix}
        m_t & \s{\wh{P}} \\
        \s{\wh{P}} & m_t
    \end{pmatrix}
    \begin{pmatrix}
        0 & |2\ra [\wh{3}| \\
        |\wh{3}]\la2| & 0
    \end{pmatrix}
    v(p_4).
\end{equation}
Again, choosing $i=3$ and $j=3$ removes the shifts on the propagator and the polarisation vector of
gluon $p_3$. However in this case all the shifts do not vanish as $|\wh{2}] = |2] - z|1]$ hence we
are left with the exact expression for the first Feynman diagram {\it plus} an extra term
coming from the surviving shifts:
\begin{equation}
    -\frac{z}{(P^2-m_t^2)\A{2}{3}^2}\bar{u}(p_1)\,
    \begin{pmatrix}
        0 & |3\ra [3| \\
        |3]\la3| & 0
    \end{pmatrix}
    \begin{pmatrix}
        m_t & \s{P} \\
        \s{P} & m_t
    \end{pmatrix}
    \begin{pmatrix}
        0 & |2\ra [3| \\
        |3]\la2| & 0
    \end{pmatrix}
    v(p_4),
\end{equation}
which simplifies to,
\begin{equation}
    -\frac{1}{\AA{2}{p_4}{3}\A{2}{3}^2}\bar{u}(p_1)\,
    \begin{pmatrix}
        0 & |3\ra\AA{2}{p_4}{3}[3| \\
        |3]\AA{3}{p_4}{3}\la2| & m_t|3]\A{3}{2}[3|
    \end{pmatrix}
    v(p_4).
    \label{eq:rec1}
\end{equation}
If the result from the recursion relation is to match the result of the Feynman calculation this
expression should be equivalent to the second diagram in figure \ref{fig:ttgg}.
Making use of the Dirac equation
one can simplify the Feynman calculation to
\begin{align}
    -\frac{1}{\A{2}{3}^2}\bar{u}(p_1)\,
    \begin{pmatrix}
        0 & p_3 \\
        p_3 & 0
    \end{pmatrix}
    v(p_4).
    \label{eq:feyn2}
\end{align}
It may not be immediately obvious that the expressions \eqref{eq:rec1} and \eqref{eq:feyn2} are
equivalent, but they are. Firstly we note that the four component spinor can be written in
terms of two component spinors,
\begin{equation}
    u(p) = \begin{pmatrix} |u_p\ra\\ |u_p] \end{pmatrix}.
\end{equation}
This allows us to expand out \eqref{eq:rec1} and immediately identify that the top row is in the correct
form. The bottom row can be simplified by using $m_t|v_4] = \s{p}_4|v_4\ra$ and by decomposing
$|v_4\ra = \alpha|3\ra+\beta|2\ra$ we can re-form the bottom row into the correct form and
reconstruct \eqref{eq:feyn2}.

This shows that recursion relations can be used to successfully calculate amplitudes
with massive particles with spin on internal lines, as expected.

%%%%%%%%%%%%%%%%%%%%%%%%%%%%%%%%%%%%%%%%%%%%%%%%%%%%%%%%%%
\section{Conclusions \label{sec:conc}}

In this paper we have applied the recursion relations derived by Britto, Cachazo, Feng
and Witten to a range of processes involving both internal and external, off-shell or
massive particles. Our analysis includes massive particles with spin and this
complements our earlier work~\cite{mrscalar} where we have considered massive scalars
coupled to massless partons. As a first step, in Sec.~\ref{sec:single} we derived
compact expressions for single vector boson currents where an off-shell gauge boson
couples to a quark-antiquark pair and up to four gluons. These results both agree with
and are more compact than the expressions previously available in the literature.  
New results for particular helicity configurations involving any number of partons
were also derived. 

Adding more off-shell vector bosons is also straightforward.   The most direct
approach is to divide the amplitude into its Abelian and Non-Abelian components.   The
Non-Abelian contribution is straightforwardly obtained by linking one or more currents
with fewer vector bosons via triple or quartic vertices. On the other hand, the
Abelian contribution is amenable to a recursive approach using the single Feynman
diagram where all of the off-shell vector bosons couple to a fermion line in an
ordered way.   As an example, Sec.~\ref{sec:double} contains compact expressions for 
the double vector boson current coupling to a quark-antiquark pair and up to two
gluons in all helicity configurations.

Note that propagating particles with mass only show up in the  Non-Abelian
contribution and we were able to organise the recursive calculation of the Abelian
part of  multi-vector boson currents  without having massive intermediate states.
However, often it is necessary to deal with propagating massive particles and the
recursion relations can handle this situation as well. The key point is to generalise
the recursion relation  to avoid using helicities for internal massive particles.  The
sum over helicities is then replaced by the conventional spin sum.  

Taken together, the generalisations of the recursion relations presented here are
capable  of providing an efficient way of  calculating tree-level amplitudes within
the Standard Model and beyond.

%%%%%%%%%%%%%%%%%%%%%%%%%%%%%%%%%%%%%%%%%%%%%%%%%%%%%%%%
\acknowledgments
We would like to thank Peter Svr\v{c}ek for useful discussions
at the outset of this work.
EWNG and VVK acknowledge the support of PPARC
through Senior Fellowships and SDB acknowledges the award of a
PPARC studentship.

%%%%%%%%%%%%%%%%%%%%%%%%%%%%%%%%%%%%%%%%%%%%%%%%%%%%

\appendix
\section{Spinor Conventions and Useful Identities \label{sec:spinor}}

We use the standard $\sigma_{\mu\,\alpha\dot{\alpha}}$ and
$\sigmab^{\dot{\alpha}\alpha}_{\mu}$ matrices to represent a vector
$p^\mu$ in spinor notation
\be
p_{\alpha\dot{\alpha}}= \, p^\mu \sigma_{\mu\,\alpha\dot{\alpha}} \ , \qquad
p^{\dot{\alpha}\alpha}= \, p^\mu \sigmab^{\dot{\alpha}\alpha}_{\mu} \ .
\ee
Spinor indices are raised and lowered with $\e$-symbols, such that
\be \label{epsids}
\sigmab_\mu^{\dot{\alpha}\alpha} =\,
\e^{\alpha\beta}\e^{\dot{\alpha}\dot{\beta}}\sigma_{\mu\,\beta\dot{\beta}}.
\ee
We also have an identity
$\sigmab_\mu^{\dot{\alpha}\alpha}\sigma^\mu_{\beta\dot{\beta}} =
2\delta_\beta^\alpha\delta_{\dot{\beta}}^{\dot{\alpha}}.$

Null vector $p_a^\mu$ is represented in terms of dotted and undotted 2-spinors as
\be
p_a^{\dot{\alpha}\alpha}= \, a^{\dot{\alpha}} a^{\alpha}.
\ee
The spinor products are defined by
\begin{align}
    \A{a}{b} &= a^\alpha b_\alpha \equiv \,
    \la a|^\alpha|b\ra_\alpha \label{angle},\\
    \B{a}{b} &= a_{\dot{\alpha}} b^{\dot\alpha} \equiv \,
    [ a|_{\dot{\alpha}}|b]^{\dot{\alpha}} .\label{square}
\end{align}
We note that the spinor summation conventions
for dotted spinors in \eqref{square} follow what is usually used in QCD literature
 and differ by a minus sign
from \cite{Witten:twstr,BCF:rec,BCFW:proof,mrscalar}.

Parity transformation requires
complex conjugation which for spinors (of a real vector) is defined as
\be
| a \ra_\alpha^{\,*} = \, |a]_{\dot{\alpha}} \quad \Rightarrow \quad
\la ab\ra^* = -\,  [ab] .
\ee

\def\a{\alpha}
\def\b{\beta}

The spinor sandwiches are defined in such a way that
the summation over adjacent indices goes from up to down
${}^\alpha {}_\alpha$ for undotted indices and
from down to up 
${}_{\dot\alpha} {}^{\dot\alpha} $ for dotted indices, precisely as in
\eqref{angle} and \eqref{square}
\bea
\langle i | p |j] &&=\,   \lambda_i^{\a}\, p_{\a\dot{\a}}\,
\tilde{\lambda}_j^{\dot{\a}} ,\label{Asix}
\\
\langle i | p_r p_s |j]&&=\, 
\lambda_i^\a \, p_{r\,\a\dot{\a}}\,\,  p_{s}^{\,\dot{\a}\b}\,
\lambda_{j\,\b},
\\ \label{1.7}
{} [ i | p_r p_s | j ] &&=\,
\tilde{\lambda}_{i\,\dot{\a}} \, p_{r}^{\dot{\a}\a}\,\,
p_{s \,\a\dot{\b}}\, \tilde{\lambda}_{j}^{\,\dot{\b}} .
\eea
For
massless momenta we have
\bea
s_{ij} = \langle i\, j\rangle [j \, i] \ , \quad &&\langle i | p_r
|j] =  \langle i\, r\rangle [r\, j] \ ,
\\
\langle i | p_r p_s| j\rangle = \langle i\, r \rangle [r\, s]
\langle s\, j\rangle \ , \quad && [ i | p_r p_s | j ] =
  [ i\, r] \langle r\, s\rangle [s\, j] \ ,
\eea
and so on.

Equation \eqref{Asix} also implies that
\be \label{Anine}
    \AA{a}{\sigma_\mu}{b} = \la a |^\alpha \sigma_{\mu\,\alpha\dot{\alpha}}
    |b]^{\dot{\alpha}}\ , \quad
    \BB{b}{\sigmab_\mu}{a} = [b|_{\dot{\alpha}} \sigmab^{\dot{\alpha}\alpha}_{\mu} | a \ra_\alpha
\ee
and the identity \eqref{epsids} gives
\begin{align}
\label{Athirteen}
    \AA{a}{\sigma_\mu}{b} &= \BB{b}{\sigmab_\mu}{a},
    \hspace{0pt}\intertext{or more generally}
    \AA{a}{\ldots \sigma_\mu \ldots}{b} &= \BB{b}{\ldots \sigmab_\mu \ldots}{a} ,\\
    \Aa{a}{\ldots \sigma_\mu \ldots}{b} &= -\Aa{b}{\ldots \sigmab_\mu \ldots}{a}, \\
    \Bb{a}{\ldots \sigma_\mu \ldots}{b} &= -\Bb{b}{\ldots \sigmab_\mu \ldots}{a}.
\end{align}

\providecommand{\href}[2]{#2}\begingroup\raggedright\endgroup

\end{document}